\documentclass[a4paper,11pt]{article}
\usepackage{pos,bm,color,wrapfig}

\newcommand{\bitem}{\begin{itemize}\item}
\newcommand{\eitem}{\end{itemize}}
\newcommand{\beq}{\begin{equation*}}
\newcommand{\eeq}{\end{equation*}}
\newcommand{\beqs}{\begin{eqnarray*}}
\newcommand{\eeqs}{\end{eqnarray*}}

\newcommand{\me}[3]{\langle #1\vert\ #2\ \vert #3\rangle}
\newcommand{\qbar}{\overline{q}}
\newcommand{\ubar}{\overline{u}}
\newcommand{\dbar}{\overline{d}}
\newcommand{\sbar}{\overline{s}}

\title{Update on Glueballs}

\author*[a]{Colin Morningstar}

\affiliation[a]{Department of Physics, Carnegie Mellon University, 
Pittsburgh, Pennsylvania 15213, USA}

\emailAdd{cmorning@andrew.cmu.edu}

\abstract{
The recent BESIII announcement of a pseudoscalar glueball candidate makes an update
on glueballs from lattice QCD timely.  A brief review of how glueballs
are studied in lattice QCD is given, and the reasons that glueballs are 
difficult to study both in lattice QCD with dynamical quarks and in experiments
are outlined.  Recent glueball studies in lattice QCD are then presented, and 
an exploratory investigation of the scalar glueball using glueball, meson,
and meson-meson operators is summarized, suggesting that no scalar state below
2~GeV or so can be considered to be predominantly a glueball state.}

\FullConference{The 41st International Symposium on Lattice Field Theory (LATTICE2024)\\
 28 July - 3 August 2024\\
Liverpool, UK\\}

\begin{document}
\maketitle

\section{Introduction}

\begin{figure}[b]
\begin{center}
\includegraphics[width=2.5in]{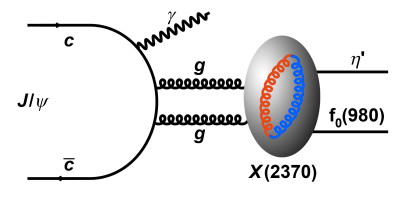}
\includegraphics[width=1.5in]{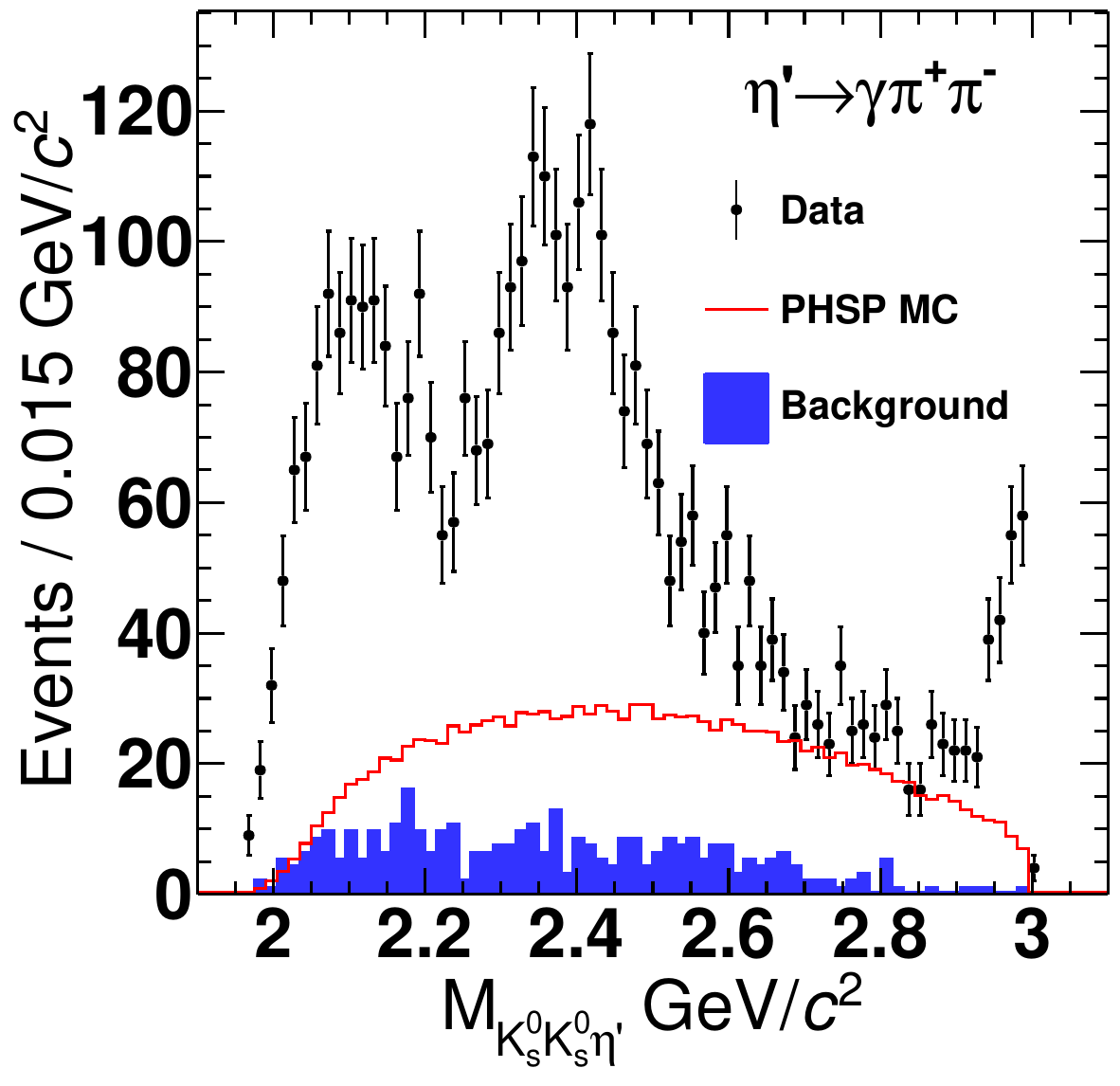}
\includegraphics[width=1.5in]{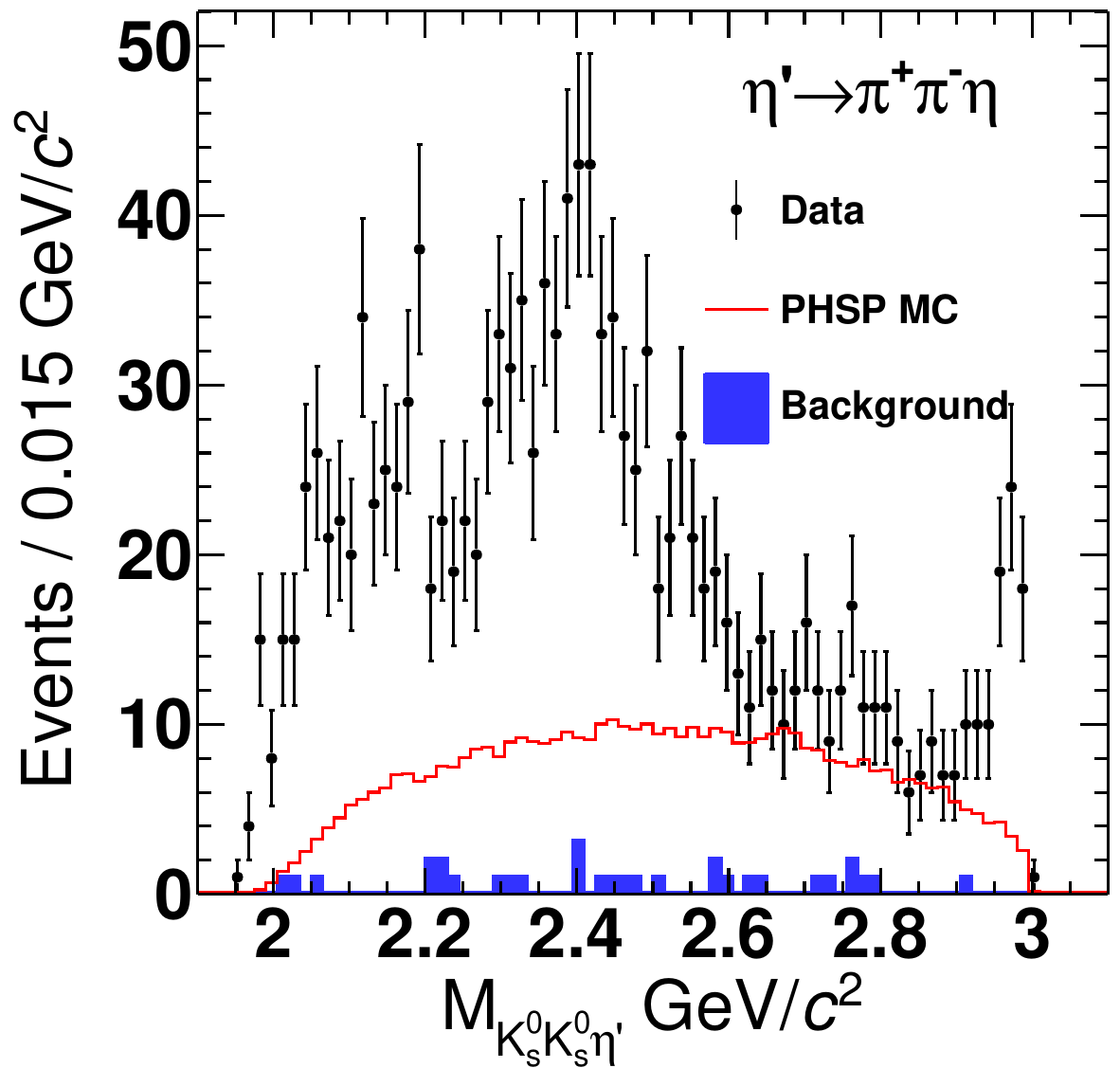}
\end{center}
\vspace*{-3mm}
\caption{Left: a typical process in BESIII which can produce the $X(2370)$.  
Right: the $K^0_S K^0_S \eta^\prime$ invariant mass distributions
with the requirement $M_{K^0_S K^0_S} < 1.1~{\rm GeV}/c^2$ for 
$\eta^\prime\rightarrow \gamma \pi^+\pi^-$ and $\eta^\prime\rightarrow \pi^+\pi^-\eta$ channels.
Data are indicated by the dots with error bars, and the shaded histograms 
are the non-$\eta^\prime$ backgrounds. The solid red lines are phase space
Monte Carlo events with arbitrary normalization. Plots are taken from 
Ref.~\cite{PhysRevLett.132.181901}.
\label{fig:X2370diag}}
\end{figure}

\begin{figure}[t]
\begin{center}
\raisebox{5mm}{\includegraphics[width=2.3in]{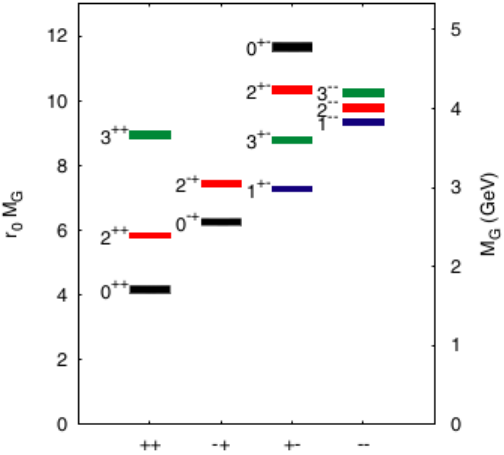}}
\hspace*{5mm}
\includegraphics[width=2.5in]{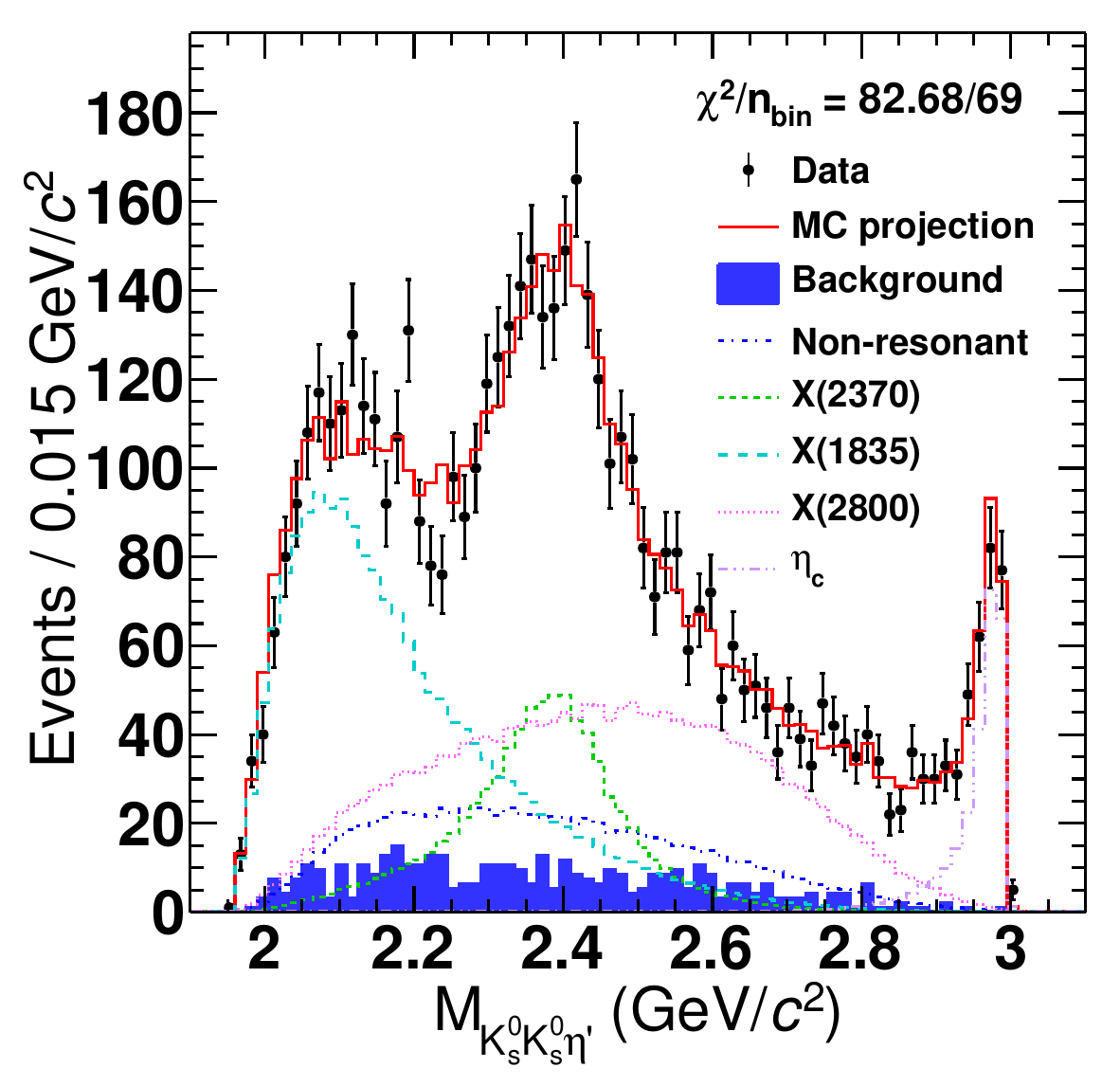}
\end{center}
\vspace*{-3mm}
\caption{Left: the mass spectrum of glueballs in the pure gauge Yang-Mills theory
from Ref.~\cite{PhysRevD.73.014516}. The masses are given both in terms of $r_0$
($r_0^{-1} = 410$~MeV) and in GeV. The height of each colored
box indicates the statistical uncertainty of the mass.
Right: a comparison between data and PWA fit projections
from Ref.~\cite{PhysRevLett.132.181901} showing the invariant mass distributions of 
$K^0_S K^0_S \eta^\prime$.  Data show as dots with error bars, and the solid red histograms 
are the PWA total projections. The shaded histograms are the non-$\eta^\prime$ backgrounds.
Contributions from the $X(2370)$, $X(1835)$, $X(2800)$, $\eta_c$, and non-resonant processes
are shown by the short dashed green, long dashed cyan, dotted magenta, dash-dot-dotted
violet, and dash-dotted blue lines, respectively.
\label{fig:pwafit}}
\end{figure}

An update on glueballs from lattice QCD is timely now due to a first-time determination 
by BESIII of the $0^{-+}$ quantum numbers of the $X(2370)$ resonance\cite{PhysRevLett.132.181901}.
This resonance, previously found in $J/\psi\rightarrow \gamma\pi^+\pi^-\eta^\prime$ and reported 
in Ref.~\cite{PhysRevLett.106.072002}, occurs in a gluon rich environment (see Fig.~\ref{fig:X2370diag})
and its mass is consistent with the lightest $0^{-+}$ glueball from pure-gauge lattice 
QCD\cite{BALI1993378,PhysRevD.60.034509,PhysRevD.73.014516,Gregory:2012hu,PhysRevD.100.054511}
(see Fig.~\ref{fig:pwafit}), 
leading some to speculate that it might be a glueball.  A partial wave analysis (PWA) of 
$J/\psi\rightarrow\gamma K^{0}_{S}K^{0}_{S}\eta^{\prime}$ gives
\begin{eqnarray}
    m &=& 2395 \pm 11 ({\rm stat})^{+26}_{-94}({\rm syst})\ \mathrm{MeV}/c^{2},\\
    \Gamma &=& 188^{+18}_{-17}({\rm stat})^{+124}_{-33}({\rm syst})~\mathrm{MeV}.
\end{eqnarray}
The optimal PWA fit (see Fig.~\ref{fig:pwafit}) contains the $X(1835), X(2370), \eta_c$ and a 
broad $0^{-+}$ $X(2800)$ Breit-Wigner with decays through $f_0(980)\eta'$ to $(K_S^0K_S^0)_S\eta'$
and $(K_S^0K_S^0)_D\eta'$ with nonresonant components, producing a statistical significance of 
the $X(2370)$ of $>11.7\sigma$.

Identifying a glueball in experiments is notoriously difficult.  First, there are no
reliable estimates of their masses from theory to help guide experimental searches.
To date, lattice QCD has only provided reliable glueball mass determinations in
the pure gauge theory without dynamical quarks.  Second, one expects flavor symmetric decays, 
but differing quark masses can lead to differing phase spaces which could affect branching ratios.
There are no rigorous predictions from theory on decay patterns and their branching ratios.
Glueball decays could be similar to that of charmonium states, and observed resonances
could be admixtures having both glueball and quark-antiquark components. Early glueball candidates
were the light scalar candidates $f_0(1370), f_0(1500), f_0(1710)$ from
MarkII in the 1980s and Crystal Barrel in the 1990s.  The narrow $\xi(2230)$ tensor glueball 
candidate from MarkIII in the 1980s and BESI in the 1990s possessed good flavor-symmetric decay
properties, but its existence was not confirmed later by BESII nor BESIII with much higher statistics.
An odderon (odd $C$-parity) from D0 and TOTEM\cite{PhysRevLett.127.062003} has also been suggested
as a glueball candidate.  Given the above considerations, an identification of the latest pseudoscalar
candidate from BESIII as a glueball cannot be considered definitive.

The pure-gauge glueball spectrum has a long history in lattice QCD, with calculations dating
back to early days of lattice QCD in the 1970s.  Given their heavy masses (the lightest is
around 1.6~GeV), the temporal correlators used to extract their masses fall off very rapidly
in imaginary time with statistical uncertainties that do not diminish significantly with time,
leading to a very quickly degrading signal-to-noise in Markov-chain Monte Carlo computations
as the temporal separation of the source and sink operators increases.  Significant progress
in calculating the spectrum of glueballs was made in the 1980s and 1990s by
M.~Teper, C.~Michael, D.~Weingarten, among others.  The introduction of anisotropic
lattices in the late 1990s allowed much better temporal resolution of the correlators,
significantly improving glueball energy determinations.  Today, the mass spectrum of
glueballs in the pure-gauge Yang-Mills theory is well known (see Fig.~\ref{fig:pwafit}).
Their mass ratios are well determined, but obtaining their masses in MeV, for example, is
less straightforward due to scale setting ambiguities.  The pure-gauge theory is not
physical (quark loops cannot be suppressed in nature), so setting any quantity computed
in the pure-gauge theory by using the value of the analogous observable obtained in 
experiment is problematic.  The string tension from the static quark-antiquark potential
or some variant of it is generally used to set the scale, leading to the lightest glueball
having a mass around $1600-1700$~MeV.  The Clay Mathematics Institute is offering a
\$1 million bounty\cite{clayYMgap} for a mathematical proof of the existence of the
mass gap of the lightest glueball, which has the quantum numbers of a scalar particle.
This Yang-Mills mass gap is one of its Millenium Prize problems.

In lattice QCD, finite-volume stationary-state energies are extracted from a matrix of
temporal correlation functions,
$C_{ij}(t) = \me{0}{O_i(t)\,\overline{O}_j(0)}{0}$,
where the set of operators $\overline{O}_i(t)$ creates the states of interest at
imaginary time $t$.  In the scalar channel, a vacuum-expectation value subtraction
is useful.  Because of the finite spatial volume and the usual imposition
of periodic boundary conditions, the energies of the stationary states are discrete.  
Neglecting temporal wrap-around effects, these correlation matrix elements have spectral
representations
of the form
\beq
   C_{ij}(t) = \sum_n Z_i^{(n)} Z_j^{(n)\ast}\ e^{-E_n t},
   \qquad\quad Z_j^{(n)}=  \me{0}{O_j}{n}.
\eeq
It is not practical to do fits using the above form due to the large number of
unknown parameters that must be determined. To extract the finite-volume energies $E_n$ 
and operator overlap factors $Z_j^{(n)}$, we define a new correlation matrix 
$\widetilde{C}(t)$ using a single pivot
\beq
   \widetilde{C}(t) = U_D^\dagger\ C(\tau_0)^{-1/2}\ C(t)\ C(\tau_0)^{-1/2}\ U_D,
\eeq
where the columns of $U_D$ are the eigenvectors of
   $C(\tau_0)^{-1/2}\,C(\tau_D)\,C(\tau_0)^{-1/2}$, with $\tau_D>\tau_0$.
We choose $\tau_0$ and $\tau_D$ large enough so that the matrix $\widetilde{C}(t)$ 
stays diagonal, within statistical errors, for $t>\tau_D$.
Then we can use single- or two-exponential fits to $\widetilde{C}_{\alpha\alpha}(t)$
to obtain the energies $E_\alpha$ and overlaps $Z_j^{(n)}$.

\begin{figure}
  \begin{center}
 \raisebox{4mm}{\includegraphics[width=3.0in]{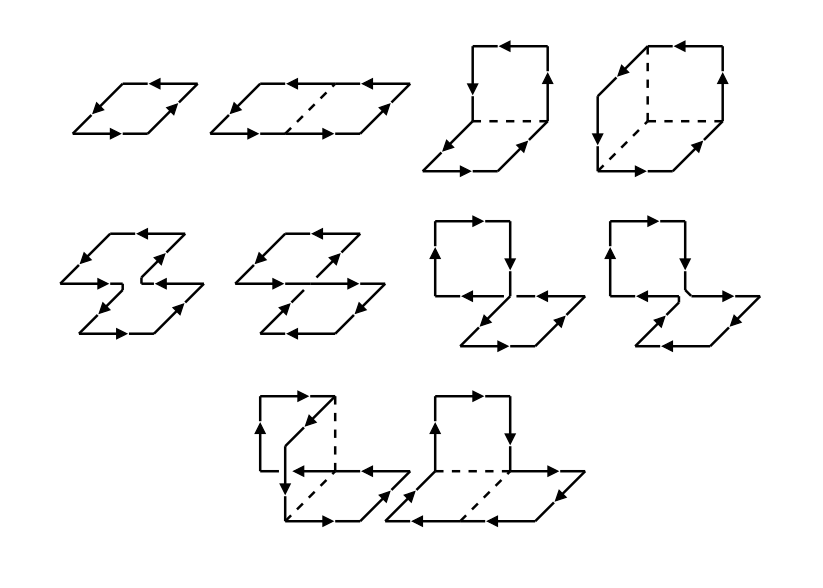}}
 \includegraphics[width=2.5in]{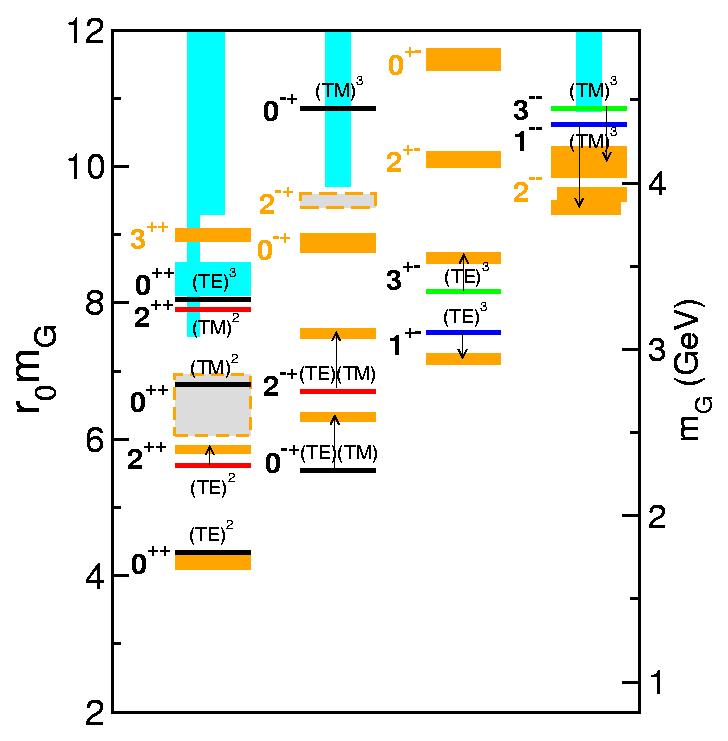}
 \end{center}
 \caption{Left: the various typical Wilson loop shapes used in making 
  glueball operators.  Each link represents a smeared path which is one
  or more lattice spacings in length.  Right: comparison of the glueball
  spectrum from the MIT bag model (with revised parameter values) with
  that from lattice QCD (shown as orange boxes).
 \label{fig:opsandbag}}
\end{figure}

Due to the unfavorable signal-to-noise of correlators involving glueball operators,
it is imperative to use very good operators to ensure reliable energy extractions
at time separations as small as possible.  Fortunately, much work has been done
in designing such operators.  First, a variety of spatial link smearings are performed,
such as Teper fuzzing\cite{Teper1987345} and stout-link smearing\cite{Morningstar:2003gk}.
In particular, Teper fuzzing creates paths which are one or more lattice links long,
allowing spatially large operators to be efficiently evaluated.  Glueball operators
are then formed from gauge-invariant loops of the smeared link variables, such as
those shown in Fig.~\ref{fig:opsandbag}.  Multiple sizes of loops with different
shapes are constructed to build up the necessary radial and orbital structures.

\begin{table}
\caption{Quantum numbers associated with the glueball operators
of minimal dimension, taken from Ref.~\cite{Jaffe1986344}
\label{tab:jaffeops}}
\vspace*{-10mm}
\begin{center}
\begin{tabular}{cll}
 Dimension & Operators & Quantum numbers \\
  4 & $ {\rm Tr}F_{\mu\nu}F_{\alpha\beta}$ & $0^{++}, 0^{-+}, 2^{++}, 2^{-+} $\\
  5 & $ {\rm Tr}F_{\mu\nu}D_\rho F_{\alpha\beta}$ & $ 1^{++}, 3^{++}$\\
  6 & ${\rm Tr}F_{\mu\nu}F_{\rho\omega}F_{\alpha\beta}$ &
        $ 0^{\pm +}, 1^{\pm\pm}, 2^{\pm\pm}, 3^{\pm -}$\\
  6 & ${\rm Tr}F_{\mu\nu}\{D_\rho,D_\omega\}F_{\alpha\beta}$ &
        $ 1^{-+}, 3^{-+}, 4^{\pm +}$
\end{tabular}
\end{center}
\end{table}

The glueball spectrum can be qualitatively understood in terms of interpolating 
operators of minimal dimension, as first outlined long ago in Ref.~\cite{Jaffe1986344}
and listed in Table~\ref{tab:jaffeops}.  Of the lightest six glueball states from
lattice QCD, four have the $J^{PC}$ of the dimension-four operators, and the
absence of low-lying $0^{\pm -},\ 1^{-+}$ glueballs is explained.
The spectrum also agrees qualitatively with the MIT bag model, as shown in
Fig.~\ref{fig:opsandbag}.  In this early model, constituent gluons are transverse
chromoelectric (TE) or transverse chromomagnetic (TM) modes in a spherical cavity,
and the spectrum is obtained from Hartree modes with residual perturbative 
interactions and a center-of-mass correction.  The results in Fig.~\ref{fig:opsandbag}
come from Ref.~\cite{PhysRevD.27.1556} with parameter modifications
suggested in Ref.~\cite{kutibagmodel}: strong coupling
$\alpha_s: 1.0\rightarrow 0.5$ and bag parameter 
$B^{1/4}: 230~{\rm MeV}\rightarrow 280~{\rm Mev}$.  The flux-tube model of
glueballs in Ref.~\cite{PhysRevD.31.2910} yields a mass spectrum in
striking disagreement with that from lattice QCD.

Studying glueballs in lattice QCD with dynamical quarks is very challenging.
To extract the energies of the heavy glueball states, the energies of all levels
lying below the glueballs of interest must also be extracted, and there are
many two-meson, three-meson, and four-meson levels expected.  Multi-meson
correlators typically require costly timeslice-to-timeslice propagators.
Glueballs are expected to be unstable resonance states whose masses and widths
must be deduced from fits of scattering $K$-matrix parametrizations to the
finite-volume spectra via a L\"{u}scher quantization condition.  Carrying out
such fits can be a daunting task.  As previously mentioned, correlators
involving glueballs are statistically noisy, so high statistics are required
which is very difficult with dynamical quarks.  Large vacuum expectation
values must also be subtracted in the scalar sector, exacerbating the
difficulties.

\section{Some recent glueball studies}
In the remainder of this talk, I highlight some recent glueball
studies in lattice QCD.

\subsection{Glueballs with \boldmath{$N_f=4$} light quarks}
\begin{figure}[t]
\begin{center}
\includegraphics[width=3.0in]{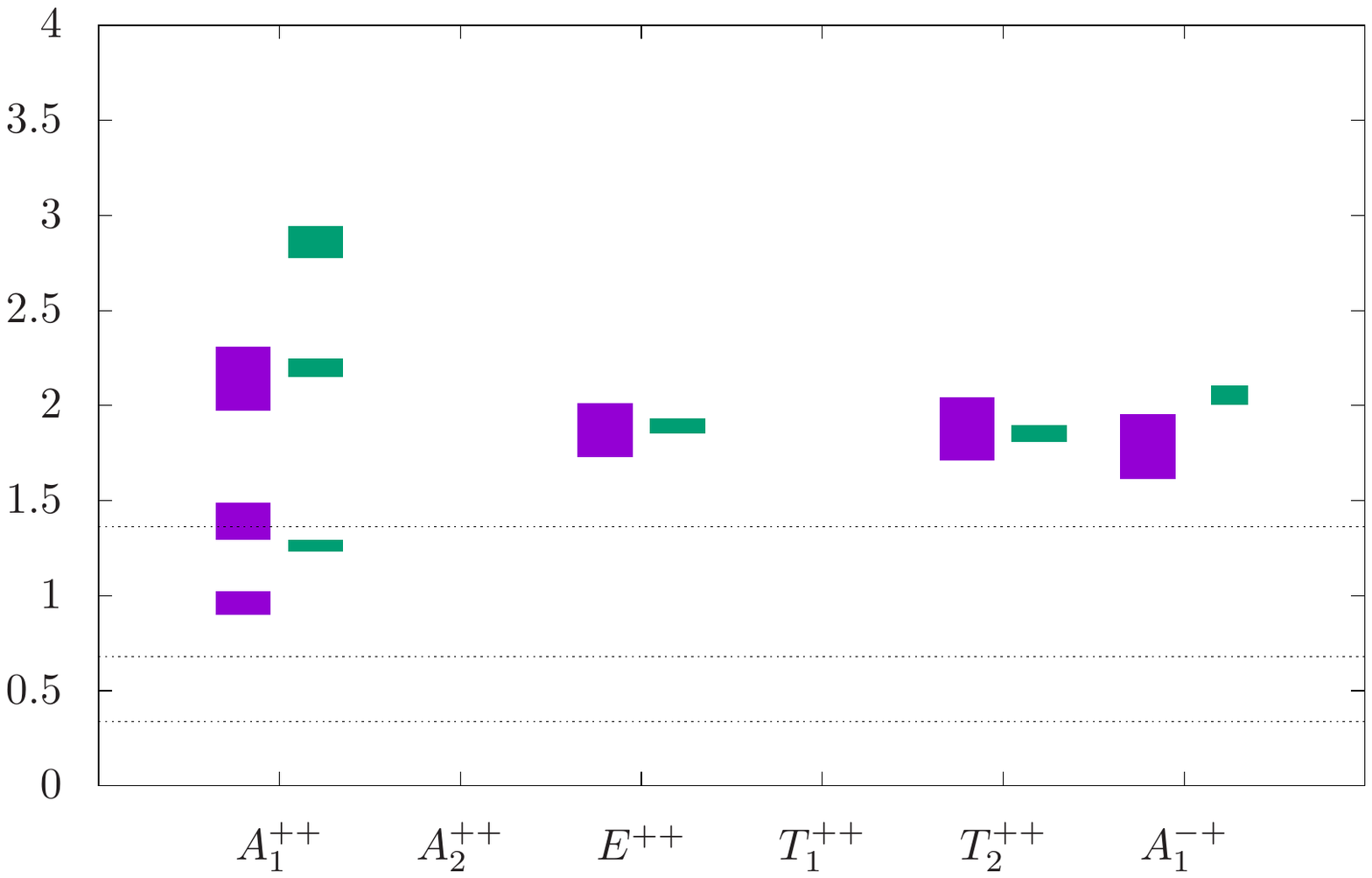}
\end{center}
\vspace*{-6mm}
\caption{The spectrum of glueballs for the representations 
$A^{++},E^{++}, T_2^{++}, A_1^{-+}$ from Ref.~\cite{Athenodorou:2023ntf}. The
vertical scale is mass times $\sqrt{t_0}$,
where $t_0$ is the usual gradient flow parameter\cite{Luscher:2010iy}.  
Results from $N_f=4$ QCD for a particular
ensemble are shown in purple, while the states in SU(3) pure gauge are shown
in green. The dashed lines correspond to 1, 2 and 4 times the pion mass from 
bottom to top, respectively. 
\label{fig:tepernf4}}
\end{figure}

A study of glueballs with $N_f=4$ light quarks has recently
been presented in Ref.~\cite{Athenodorou:2023ntf} with the goal of examining
the effect of quark loops on the glueball spectrum.  Several ensembles were
used with $m_\pi\approx 250~{\rm MeV}$.  The energies obtained are compared
to the pure gauge theory in Fig.~\ref{fig:tepernf4}. In this figure, one
sees that the scalar results are lowered towards the $2\pi$ threshold,
and the tensor and pseudoscalar masses are only slightly affected.
This work is exploratory since only glueball operators were used.  
No meson-meson operators were incorporated into the study.  The topological
charge is evaluated, and the string tension is found to be suppressed by the 
inclusion of four light dynamical quarks by 50-70\%, compared to the
pure gauge theory.

\begin{figure}[b]
\begin{center}
\includegraphics[width=2.5in]{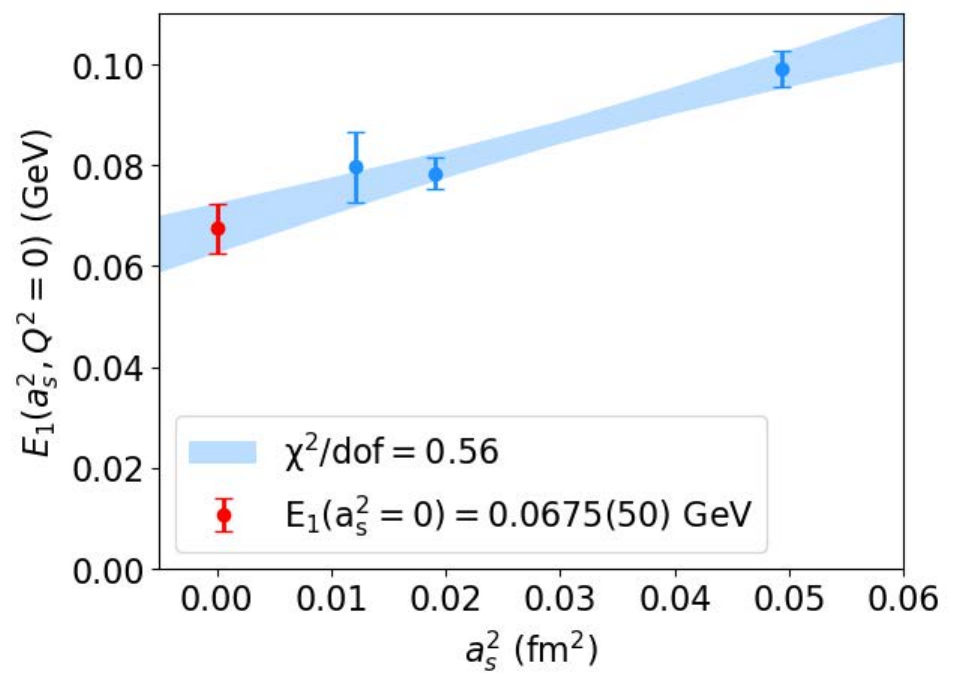}\qquad
\includegraphics[width=2.45in]{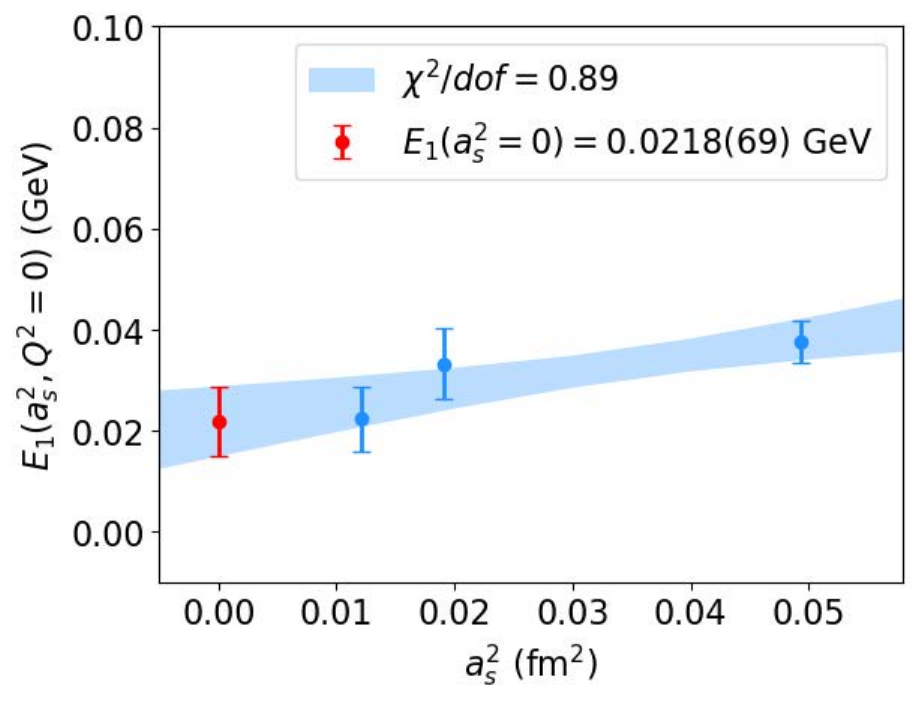}
\end{center}
\vspace*{-6mm}
\caption{Linear continuum limit extrapolations of the $E_1$ form factors
for the $J/\psi \rightarrow\gamma G$ process (left) and $J/\psi \rightarrow\gamma \phi$ 
(right) using three different lattice spacings from Ref.~\cite{Zou:2024ksc}.
\label{fig:zou}}
\end{figure}

\subsection{Radiative decay of the scalar glueball}
The radiative decay of the scalar glueball was studied in the quenched
approximation in Ref.~\cite{Zou:2024ksc} using three gauge ensembles
with lattice spacings $a_s\sim 0.11, 0.14, 0.22~{\rm fm}$ and extrapolating
to the continuum limit.  The EM transition matrix element 
$\langle S\vert J^\mu_{\rm em}\vert V\rangle$ was evaluated and a
multipole expansion used to obtain two form factors $E_1(Q^2)$ and $C_1(Q^2)$.
Decay widths are obtained from $E_1(0)$ using $Q^2\rightarrow 0,\ a\rightarrow 0$
extrapolations.  The continuum limit extrapolations for $E_1(0)$ for
the $J/\psi \rightarrow\gamma G$ and $J/\psi \rightarrow\gamma \phi$ 
processes are shown in Fig.~\ref{fig:zou}.  This work finds
$\Gamma(J/\psi\rightarrow \gamma G) = 0.578(86)~{\rm keV}$
with ${\rm Br}(J/\psi\rightarrow\gamma G)=6.2(9)\times 10^{-3}$
and $\Gamma(G\rightarrow\gamma\phi) = 0.074(47)~{\rm keV}$,
concluding that $J\psi\rightarrow\gamma G\rightarrow\gamma\gamma\phi$
is not detectable by BESIII.

\begin{figure}[t]
\begin{center}
\includegraphics[width=5.0in]{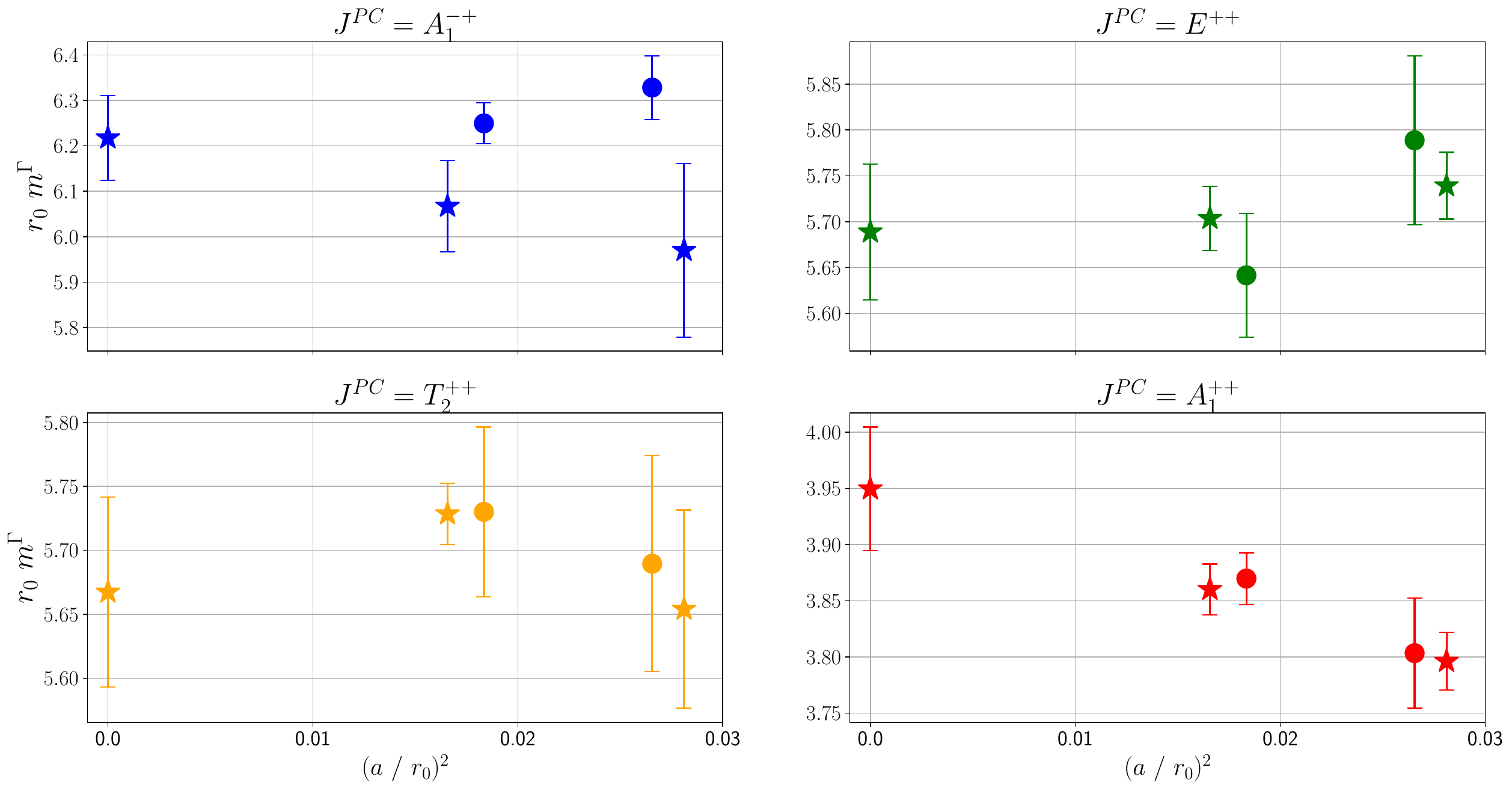}
\end{center}
\vspace*{-6mm}
\caption{Comparison between Wilson-action glueball masses obtained in 
Ref.~\cite{Barca:2024fpc} at $\beta = 6.2, 6.08$ using a new multi-level 
error reduction algorithm (circles) and state-of-the-art results (stars) 
at $\beta = 6.0625, 6.235$ and in the continuum limit from a recent 
study\cite{Athenodorou:2020ani}
using the traditional Monte Carlo algorithm.
\label{fig:errreduce}}
\end{figure}

\subsection{Error reduction algorithm}
In Ref.~\cite{Barca:2024fpc}, a new multi-level sampling procedure 
was proposed for error reduction of glueball correlators in the pure $SU(3)$
gauge theory. Comparisons of some glueball masses obtained using this new 
method with masses obtained using the traditional 
procedure\cite{Athenodorou:2020ani} are shown in 
Fig.~\ref{fig:errreduce} for the Wilson action.  Although no significant 
reduction in the glueball 
mass errors were observed, significant error reduction in large$-t$ correlators
were found, which improves confidence in plateau estimates.

\begin{figure}[b]
\begin{center}
\includegraphics[width=2.8in]{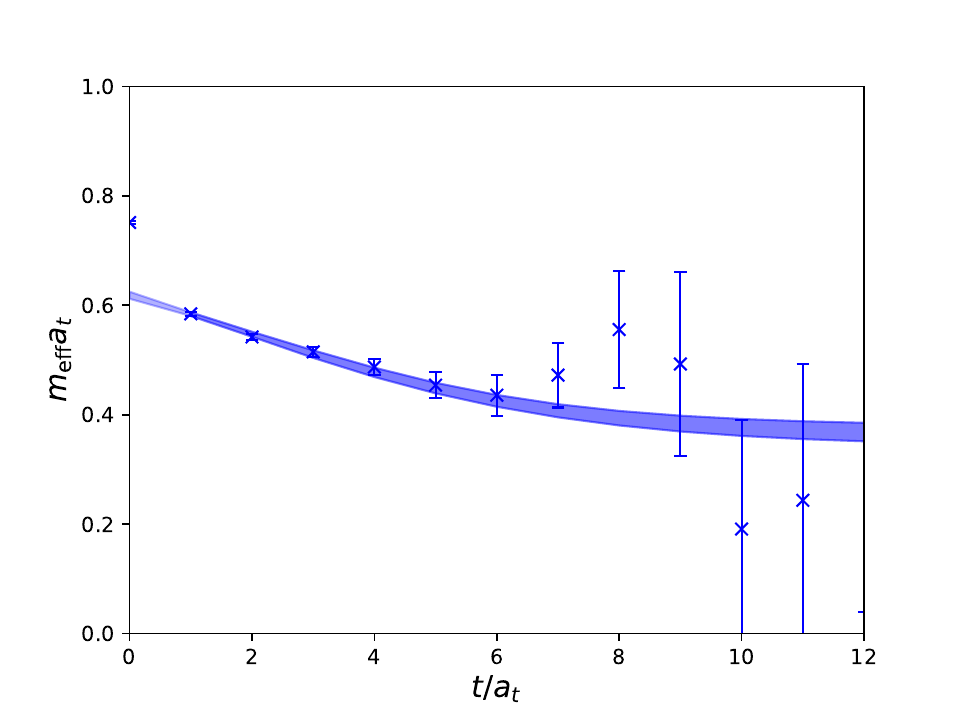}
\includegraphics[width=2.8in]{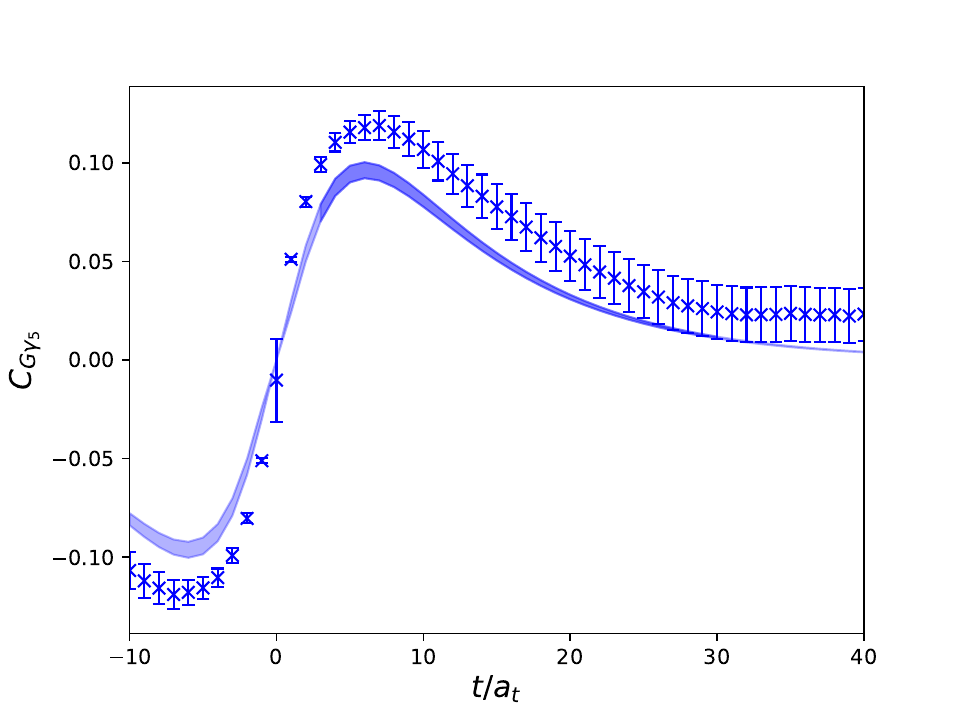}
\end{center}
\vspace*{-5mm}
\caption{(Left) Effective mass for the $0^{-+}$ glueball correlator from
Ref.~\cite{Jiang:2022ffl}. The shaded band shows the result from a forward and 
backward two-exponential fit. (Right) The glueball-$\eta$ cross correlator 
(shifted horizontally).  The blue band shows the result from a fit to the 
temporal derivative of the correlator using forward-backward exponentials
with four masses.
\label{fig:getamix}}
\end{figure}

\subsection{Glueball-\boldmath{$\eta$} mixing}
The mixing of the $0^{-+}$ glueball and the pseudoscalar $\eta$ meson
was studied in Ref.~\cite{Jiang:2022ffl}, with results shown in 
Fig.~\ref{fig:getamix}. Results were obtained on a $16^3\times 128$ 
anisotropic $N_f=2$ lattice with $m_\pi\approx 350~{\rm MeV}$.  The distillation 
method\cite{Distillation:2009krc} was utilized.  The pseudoscalar glueball
operator was constructed using a variationally optimized superposition of a 
variety of smeared gauge-invariant loops, as usual.  The $\eta$ operator
used was a local $\overline{q}\gamma_5 q$ operator using smeared quark fields,
although $\overline{q}\gamma_4\gamma_5 q$ was also considered.  The diagonal
and cross correlators were computed, and a very small $3.5^\circ$ mixing angle 
was obtained from the glueball-$\eta$ cross correlator.

\begin{figure}[t]
\begin{center}
\includegraphics[width=5.5in]{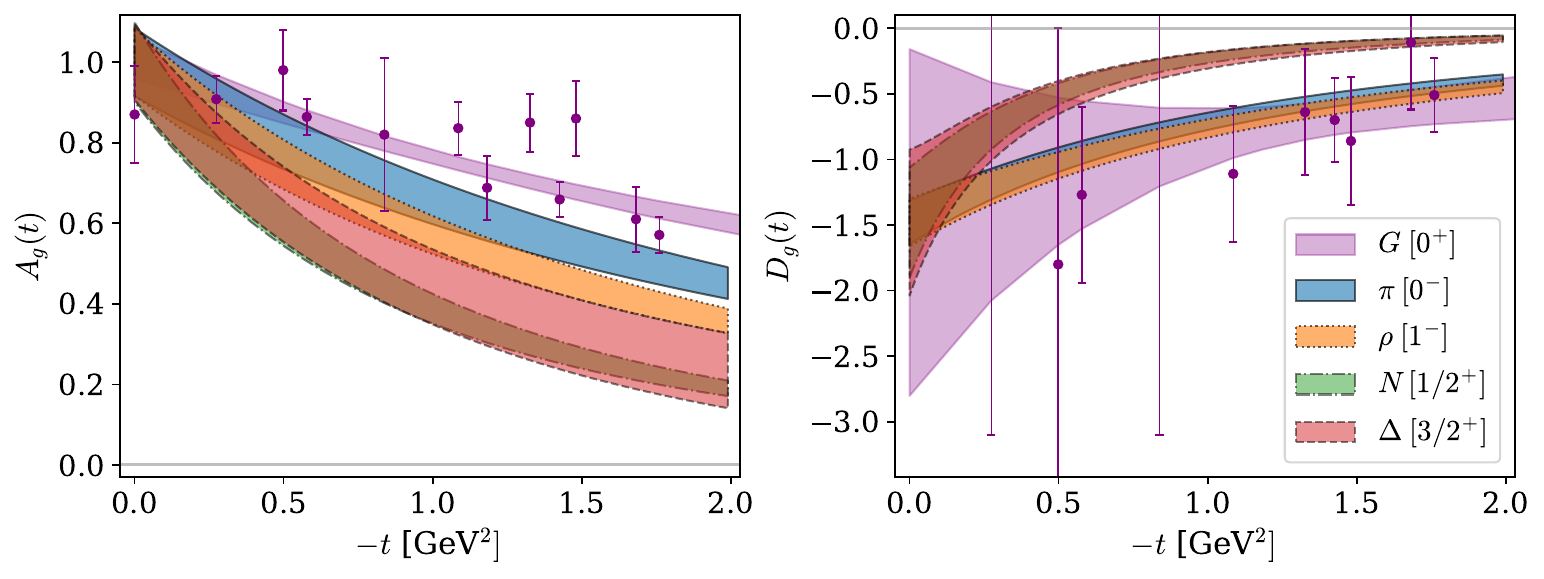}
\end{center}
\vspace*{-7mm}
\caption{Comparison between the scalar glueball GFFs in Yang-Mills theory obtained 
in Ref.~\cite{Abbott:2024bre} and the gluon GFFs from Ref.~\cite{Pefkou:2021fni} of the pion, 
$\rho$ meson, nucleon, and $\Delta$ baryon, obtained with an $N_f = 2 + 1$ QCD 
ensemble with $m_\pi = 450$~MeV. As usual, $t=(p^\prime-p)^2$, where $p$ and $p^\prime$
are the four-momenta of the incoming and outgoing states.
\label{fig:GFFs}}
\end{figure}

\subsection{Gravitational form factors of glueballs}
In Ref.~\cite{Abbott:2024bre}, a first step towards probing the structure of
glueballs using gravitational form factors (GFF) was presented at this conference.
The GFFs are obtained from the energy-momentum tensor matrix elements in 
SU(3) pure gauge theory.  The matrix element of $T_{\mu\nu}$ in the scalar 
glueball state can be expressed in terms of two form factors $A(t)$ and $D(t)$,
with $t=(p^\prime-p)^2$ and where $p$ and $p^\prime$ are the four-momenta of 
the incoming and outgoing states. Preliminary results in pure gauge theory on
\begin{wrapfigure}{r}{0.5\textwidth}
\begin{center}
\includegraphics[width=0.45\textwidth]{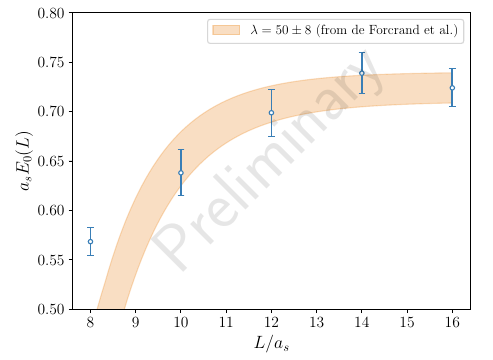}
\end{center}
\vspace*{-6mm}
\caption{The $A_1^{++}$ energy in pure Yang-Mills in finite volume $L^3$
from Ref.~\cite{HansenGB2024} against $L$.  The trilinear coupling 
$\lambda$ was then extracted by a fit to the L\"{u}scher relation.
\label{fig:gluonscat}}
\end{wrapfigure}
a $24^3\times 48$ lattice with spacing $a=0.1~{\rm fm}$ were presented and
are shown in Fig.~\ref{fig:GFFs}.  
The glueball GFFs in this figure are compared
to gluon GFFs from Ref.~\cite{Pefkou:2021fni} of the pion, 
$\rho$ meson, nucleon, and $\Delta$ baryon, obtained with an $N_f = 2 + 1$ QCD 
ensemble with $m_\pi = 450$~MeV.

\subsection{Scalar glueball scattering}
First steps towards computing glueball-glueball scattering in lattice
QCD were presented in Ref.~\cite{HansenGB2024}.  In this study, 
finite-volume energies in Yang-Mills involving both single and two scalar
glueball operators were computed. An anisotropic lattice was used, a multi-level 
algorithm was employed to reduce correlator errors, and scale setting was done 
using the gradient flow parameter $t_0$ \cite{Luscher:2010iy}.  The volume 
dependence of the
$A_1^{++}$ energy was evaluated and is shown in Fig.~\ref{fig:gluonscat}.
The L\"{u}scher relation was then utilized to get the trilinear coupling 
$\lambda$ from these energies.

\begin{figure}
\begin{center}
 \includegraphics[width=\textwidth]{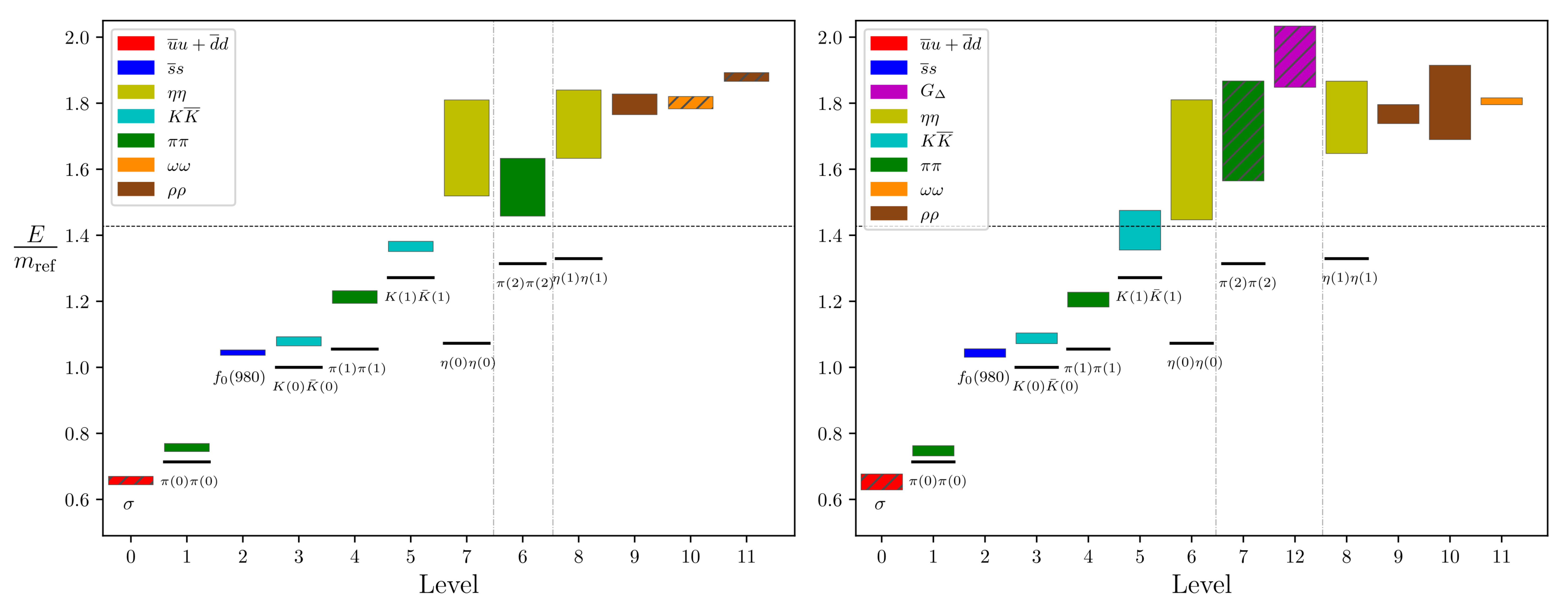}
\end{center}
\caption{Finite-volume stationary state energies in the $I=0$, $S=0$, $A_{1g}^+$
channel extracted using a $12\times12$ correlation matrix, \textit{excluding}
the scalar glueball operator on the left, and using a $13\times13$ correlation
matrix \textit{including} the scalar glueball operator on the right.
$1\sigma$ uncertainties are denoted by the box
heights. If a level is created predominantly by a single operator, the level
is colored to indicate the flavor content of that operator. If a level is
created predominantly by more than one operator, a hatched box is used to
denote the presence of operator overlaps within $75\%$ of the maximum,
indicating significant mixing. Level numbers indicate order in terms
of increasing mean energy, but the levels have been rearranged horizontally
to highlight the area of interest involving the glueball operator.
Short black
lines indicate the non-interacting two-hadron levels, and the dashed
horizontal black line indicates the $4\pi$ threshold, and
$m_{\rm ref} = 2m_K$.
\label{fig:scalarglueball}}
\end{figure}

\subsection{Scalar glueball in \boldmath{$N_f=2+1$} QCD}
The scalar glueball in the presence of dynamical quarks has been studied
previously in Refs.~\cite{Brett:2019tzr,BrettThesis}.  Although this work was
reported back in 2019, its incorporation of glueball, single quark-antiquark
meson, and meson-meson operators makes it still of interest at present.
This work was done on a small $24^3\times128$ anisotropic lattice with
$m_\pi\sim390$~MeV using the stochastic LapH method\cite{Morningstar:2011ka}.  
The main goal in this study was to discover if any 
finite-volume states below $2m_{\rm ref}$ in the vacuum sector are 
\textit{missed} when no glueball operators are included, where 
$m_{\rm ref}=2m_K$.  The results are summarized in Fig.~\ref{fig:scalarglueball}.
In the left plot of this figure, the finite-volume spectrum determined using a
basis of interpolating operators \textit{excluding} the scalar glueball operator
is shown.  A two-hadron (meson-meson) operator for each expected
non-interacting level is included, and additional operators with various flavor, 
spin, and orbital structure are added until no new finite-volume levels were 
found below $\sim 2m_{\rm ref}$.  Single-hadron $\qbar q$ operators are chosen 
in a similar way, including one of each isoscalar flavor structure: 
($\ubar u + \dbar d$, $\sbar s$) with various spatial
displacements until no new states are seen in the energy region of interest.
In the end, two $\overline{q}q$ operators and ten two-meson operators
were included to produce a $12\times 12$ correlation matrix.  In the right
plot of Fig.~\ref{fig:scalarglueball}, the spectrum obtained using a $13\times 13$
correlation matrix including the twelve operators discussed above plus
a scalar glueball operator is shown.  The low-lying spectrum is essentially
unchanged.  An additional level, shown as the purple hatched box, appears
very high in the spectrum.  Overlap factors associated with each operator
are used to identify the energy levels in Fig.~\ref{fig:scalarglueball}.
These results indicate that no finite-volume energy eigenstate below 
$\sim1.9m_{\rm ref}$ can be identified as being predominantly
created by a scalar glueball operator. As the single new energy
occurs above the region where our operator set is designed to create states, 
no reliable inferences can be drawn about this state.

While these finite-volume results are insufficient to make any definitive
statements regarding the infinite-volume resonances in this channel, we can make some
qualitative comparisons to experiment. In finding only two $\qbar q$ dominated
states below $2m_{\rm ref}$, we have observed no clearly identifiable counterpart
finite-volume $\qbar q$ states to the $f_0(1370)$, $f_0(1500)$, or $f_0(1710)$ resonances
in this region. This suggests that 
these resonances are molecular in nature rather than conventional $\qbar q$ or
pure glueball states.

\section{Conclusion}
In this talk, some recent glueball studies were highlighted.
Glueballs are very challenging to study in lattice QCD with dynamical quarks,
but progress is being made. One recent study suggests that no scalar state below 
2 GeV is predominantly a glueball state. Due to a first-time determination 
by BESIII of the $0^{-+}$ quantum numbers of the $X(2370)$ resonance,
the pseudoscalar glueball is a new focus of attention.  Identifying glueballs in
experiments is very challenging too. Glueballs have a long history in lattice QCD,
and their interesting features and the challenges of studying them ensure
they will have a long future as well.  I acknowledge support from the U.S. NSF 
under award PHY-2209167.

\bibliographystyle{JHEP}
\bibliography{references}

\end{document}